\documentstyle[sprocl,epsfig,psfig]{article}

\def\be{\begin{equation}}
\def\ee{\end{equation}}
\def\bea{\begin{eqnarray}}
\def\eea{\end{eqnarray}}

\begin{document}
\begin{flushright}
\end{flushright}
\title{HOW TO TEST THE EXISTENCE OF THE EARLY PARTON CASCADE
       USING PHOTON HBT CORRELATIONS?
\footnote{To the memory of Klaus Geiger}
\footnote{
Presented at the 2$^{nd}$ Catania Relativistic Ion Studies, CRIS'98,
Catania, Sicily, Italy, June 8-12, 1998. To appear in the 
Proceedings, Worl Scientific.
}}

\author{DANIEL FERENC}

\address{University of Regensburg,\\Physics Dep.,\\93040-Regensburg,\\
Germany\\}
\address{Div. PPE,\\CERN,\\ 1211-Geneva,\\Switzerland\\
E-mail: Daniel.Ferenc@CERN.ch} 


\maketitle\abstracts{ 
We report on a possible application
of the HBT phenomenon in testing the 
existence
of two hypothetical phenomena.
First, it is argued that the existence of a rapidly developing
parton cascade in the earliest stages of a
high energy nuclear collision process
can be tested by studying two--photon HBT correlations
over a {\it wide } 
longitudinal momentum scale -- corresponding to 
the {\it early} photon emission time from the 
hypothetical parton system.
This method provides the 
selectivity for the early emitted photons, since the 
photons emitted at later times correlate over 
progressively narrower momentum scales. 
Second, in a similar way we argue that the existence of 
a hypothetic dark matter candidate, the Weakly Interacting
Massive Particle (WIMP), may be tested 
by studying HBT correlations of cosmic gamma rays
at a relatively {\it long} detection time scale -- corresponding to 
the very {\it narrow} spectral line of the photons emerging
from WIMP annihilations. Background photons 
leave no signature since they do not correlate.
}

\section{Introduction}

The discovery of the photon intensity interference phenomenon
by
Hanbury Brown and Twiss (HBT) 
has initiated, in a rather dramatic way, 
a new field of physics -- 
the field of quantum optics~\cite{HBT-Nature}.
The phenomenon has ever since been also
utilized in different fields of science as a unique tool
capable of revealing important 
properties of various physical systems.
While in astronomy the ``HBT tool" has been 
originally used to study
angular sizes of stellar objects,
in high energy physics it has provided
information on the space-time evolution of expanding
hadronic systems created in particle collisions. 

In this paper we report on another possible application
of the HBT phenomenon, namely in testing the 
existence
of certain hypothetical phenomena.
Two applications are proposed.
First,
the study of the existence of a rapidly developing parton
cascade
in the earliest stages 
of a nuclear collision process at very high 
collision energies (at RHIC and LHC).
According to 
parton cascade models~\cite{parton-cascade}
a rapid partial parton thermalisation
establishes already
within a fraction of one fm/c time.
Very strong photon radiation -- a flash~\cite{Photon-flash} --
should be among 
the the most prominent consequences
and signatures
of the phenomenon.
In practice,
the problem is that the photons are radiated also
during all the subsequent evolution stages of the system, most
of them from the decays of neutral pions 
and other hadrons, and 
in principle there is no way to 
distinguish between the interesting early photons 
and all other photons, as long as single--photon
momentum spectra are studied. 
In that case,
the most one can do 
is to estimate the {\it time-integrated} yield of 
{\it all} the direct photons,
by subtracting 
the estimated yield of photons 
emerging from hadron decays
from the total measured contribution.

However, it is possible to distinguish the early
photon component from the rest by exploring the 
emission time information which is imprinted into the
HBT correlation pattern~\cite{photon-HBT}.
The early emitted photons
are
HBT correlated
over a very {\it wide longitudinal momentum scale}
-- thanks to their very {\it early
emission time}. 
The HBT effect appears, roughly speaking, when
the product of these two quantities becomes 
as small as $\hbar$, or in other words
when the photons are found to occupy the same 
phase space cell in the phase space (in fact in 
all three directions simultaneously).
For example, photons radiated at 
a time of 0.5 fm/c after collision will correlate over longitudinal
momentum difference of about 0.4 GeV/c.
All the other photons that emerge later are 
correlated at progressively narrower longitudinal momentum scales,
which provides a possibility to distinguish the 
contribution of the early photons, and to 
verify the existence
of the hypothetical rapidly developing parton cascade.

Another example, discussed here only briefly and
qualitatively, is the search for 
the 
Weakly Interacting
Massive Particle (WIMP, possibly the lightest SUSY particle) 
as a dark matter candidate. 
Abundant annihilation of WIMPs
into pairs of photons
may be
expected in the vicinity of very massive
cosmic objects that would assure 
high annihilation rates due to high dark matter densities.
Again, the annihilation photons will be detected together
with photons originating from
all the other sources, and
the phenomenon will be hard to 
detect,
unless a very strong annihilation
peak existed above the background (at an energy
equal to the WIMP's mass, presently expected at around
100 GeV -- in the so far unexplored energy region). 
The HBT correlation tool can in principle again provide 
selectivity, because
photons emerging from the annihilations could have a very 
narrow frequency spectrum around the WIMP mass, the spectral
width being determined by Doppler smearing due to 
WIMPs' motion, which is probably very slow.
Due to the narrow frequency spectrum,
photons from WIMP annihilations will be HBT correlated over
a very long time, while the background photons will
not be HBT correlated at all. 
A possible way to search for the annihilation process
is therefore 
by searching for the HBT temporal photon 
correlations
over a relatively long time scale. This time scale depends
on the actual WIMP's kinetic energies. In case that
the effective temperature of WIMPs would be as high as the 
microwave background temperature ($\sim$ 3 K), 
the spectral width would be $\sim 10^{-4}$ eV,
and the HBT coherence time would be $\sim 10^{-11}$ seconds. 
One of the lessons we have learned from the original
table-top photon HBT experiment~\cite{HBT-Nature} is that
the actual time-resolution of the apparatus may be
by several orders of magnitude larger than the coherence time.
Taking care of the other coordinates in phase space,
one also has to make sure that the spread in transverse 
momentum of the detected photons be sufficiently
small (small angular acceptance) to fit a 
single phase space cell.

The paper is structured as follows:
the HBT measurement tool is discussed in Sect.2.,
momentum variables are defined in Sect.3, photon
detection techniques are discussed in Sect.4,
and the results of our simple considerations are presented
and discussed in Set.5.

\section{The HBT Measurement Tool}

Statistical fluctuations of a chaotic system consisting of
noninteracting identical bosons
in the six--dimensional phase space
are not Poissonian, like for macroscopic
particles, but
of a Bose--Einstein type~\cite{Loudon}.
When a phase--space volume smaller than a single phase--space cell
is considered,
boson multiplicities fluctuate
according to the so called
geometric count probability function

\begin{equation}
P(n)=\frac{<n>^n}{(<n>+1)^{n+1}}
\end{equation}

\noindent
where $<n>$ is the average count number.
This function
is very wide --
a small number of counts is most probable,
but there is also a rather high probability to count
a large number of bosons.
The count probability distribution for a large phase--space
volume which contains many phase--space cells, approaches the
Poissonian law,
but {\it never} reaches the Poissonian limit -- 
fluctuations within single phase space
cells always remain geometric and only get diluted in bulk
as the number of independent cells increases.

The general idea of all HBT measurements
is to scan the statistical behavior of identical bosons
as a function of the size of the considered phase--space volume.
Typically one or more dimensions of the phase--space
volume are unknown, while the others are accessible to the
measurement; in particle and nuclear physics
the volume in the configuration space is unknown
(i.e. the physical size of the system at the freezeout), while the
momentum space for pions is accessible in the measurement.
The goal of a correlation measurement
is to find the scale in the accessible coordinate of the
phase--space for which the fluctuation pattern
turns from the Poissonian to the geometric law --
this indicates that the overall observed phase--space volume
(including the unobservable dimensions) corresponds to
a single phase--space cell. Knowing the volume of a 
single phase--space cell,
($\Delta p_x \Delta x \sim \hbar$, 
$\Delta p_y \Delta y \sim \hbar$,
$\Delta p_z \Delta z \sim \hbar$),
it is straightforward
to deduce the scale of the system along the unaccessible
coordinates. 

In contrast to the particle and nuclear physics, in the
famous Hanbury--Brown Twiss stellar interferometry~\cite{HBT-Nature}
the volume in the configuration space is well known and in fact
defined by the detector acceptance, while the
transverse momentum spread of light coming from a star
is the unknown required quantity; this momentum  spread $\Delta p_T$
is related to the
opening angle of the star $\theta$ as viwed from
the Earth $\Delta p_T \simeq p \theta$,
which enables one to estimate the so called angular size
of a star (not to be confused with the real size!).
In the particle and nuclear physics the
phase--space cell is placed at
the particle source, while in the stellar interferometry
it is placed between the photon detectors on the Earth.

The following methodological problem is how to check the fluctuation
pattern in real experiments. There are basically two
different methods. First, one can perform a direct
study of the count probability statistics as a function
of the observational phase--space volume; a still better
way is to study moments, e.g. factorial moments.
Another method
is to study the two--particle
correlation functions. Essentially, this means to study a
distribution of distances between particles in the accessible
part of the phase--space (in particle physics in the momentum--space
and in astronomy in configuration space)
and to compare this distribution to (or in fact divide by) a
somehow synthesized distribution
corresponding (locally) to a Poissonian statistics. 
At a certain
small scale a correlation peak emerges, indicating
where the actual statistics deviates from the Poissonian,
and thus indicating the scale corresponding to the
single phase--space cell. As long as two--particle correlations are
studied, the correlation technique is very suitable, but
when higher--order correlations are to be studied,
a serious conceptual problem arises with the definition of the
fundamental quantity -- the distance between several
particles -- and it is more natural to study
count probability moments. 

For expanding particle sources,
a correlation arises between momenta and emission coordinates
of particles.
Due to the Doppler shift,
particles
emitted from distant parts of the source cannot occupy
the same phase-space cell ($\Delta x \Delta p_x >> \hbar$)
and they are not mutually
correlated.
Only particles emitted
from a smaller relative distance $\Delta x$, and consequently
with a smaller $\Delta p_x$, demonstrate correlation. This in turn
enables us to exploit the Doppler effect in order to study
the way how the system expands.

\section{Momentum Variables and the Correlation Function}

The two single--particle momentum vectors ${\bf p}_1$ and ${\bf p}_2$
may be decomposed into the
average ${\bf k}=0.5~({\bf p}_1+{\bf p}_2)$ and 
the relative ${\bf Q}=({\bf p}_1-{\bf p}_2)$
momentum vectors of the pair,
each carrying three degrees of freedom.
The relative momentum vector may be decomposed into
transverse sideward $Q_S$, transverse outward $Q_O$,
and longitudinal $Q_L$ components ;
the longitudinal component is parallel to the collision axis,
while the transverse sideward and outward components are perpendicular
and parallel to the average transverse
momentum vector {\bf k}$_T$, respectively:

\begin{eqnarray}
{\bf Q}_L=({\bf p}_{L1}-{\bf p}_{L2})   \nonumber   \\
{\bf k}_T=\frac {{\bf p}_{T1}+{\bf p}_{T2}}{2} \nonumber   \\
{\bf Q}_T=({\bf p}_{T1}-{\bf p}_{T2})              \label{eq:11} \\
Q_S=\frac {|{\bf Q}_T\times {\bf k}_T|}{|{\bf k}_T|}   \nonumber  \\
Q_O=\frac {{\bf Q}_T\cdot {\bf k}_T}{|{\bf k}_T|}        \nonumber
\end{eqnarray}

The correlation function used in the simulations  presented
in this paper is a function of
the three relative momentum components
and was assumed to be a Gaussian in all three dimensions:

\begin{equation}
C(Q_{S},Q_{O},Q_{L})=1+\lambda e^{(-Q_{S}^2 R_{S}^2)}
e^{(-Q_{O}^2 R_{O}^2)}
e^{(-Q_{L}^2 R_{L}^2)},   
\end{equation}

\noindent
where $\lambda$ is the correlation intensity (0.5 for photons),
and $R_{S}$, $R_{O}$ and $R_{L}$ are the
parameters which characterize the photon source, or the effective interfero
metric
source sizes.

\section{Photon Detection Techniques}

The first problem to be discussed is whether and how the 
high energy
photons may be detected in a realistic experiment.
In our discussion we shall consider the ALICE 
experiment~\cite{ALICE}
at LHC/CERN. 

The experimental parameters of the highest relevance
for the 
proposed correlation study are the acceptance of the 
photon detector, and the photon detection probability.

\begin{figure}[htp]
\epsfig{file=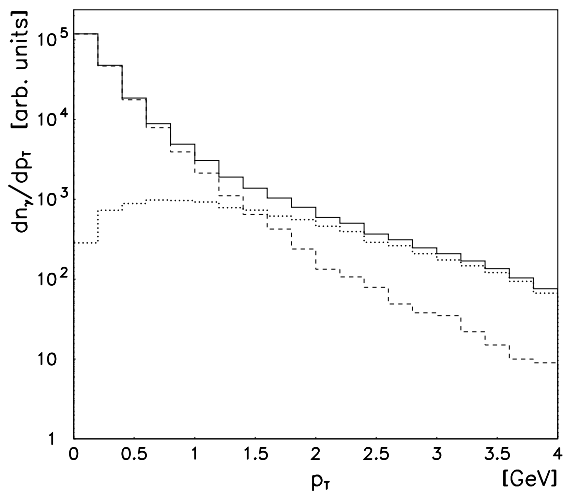,bbllx=0,bblly=0,bburx=792,bbury=750,width=7cm}
\caption{
\noindent
Transverse momentum spectra $dn/dp_T$ for the ``early photon component"
(dotted line), ``late photon component" (dashed line)
and the sum of the two (full line).
}
\label{FPP01}
\end{figure}

\begin{figure}[htp]
\epsfig{file=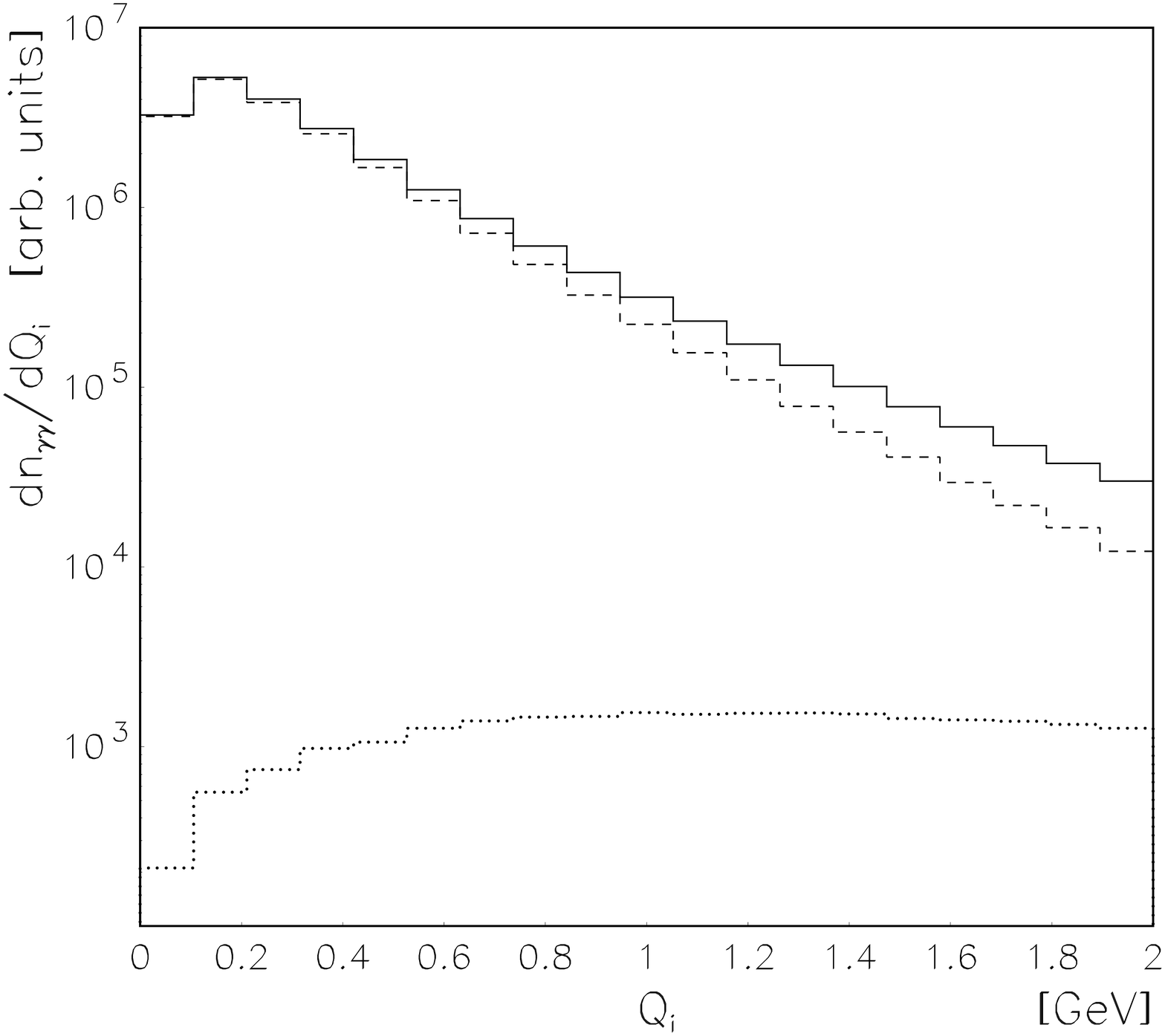,bbllx=0,bblly=0,bburx=792,bbury=480,width=7cm}
\caption{
\noindent
Distribution of photon pairs as a function of $Q_i$.
Pairs of early photons (dotted line) constitute a very small
fraction of 
the total photon contribution (full line); 
late photons (dashed line).
No $p_T$ cuts are applied.
}
\label{FPP02}
\end{figure}
\begin{figure}[hbp]
\epsfig{file=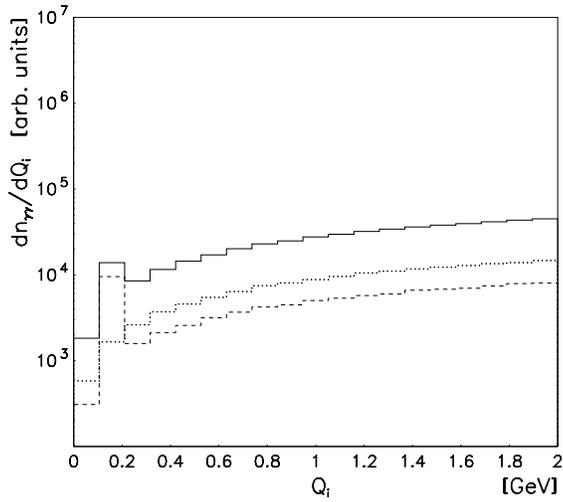,bbllx=0,bblly=0,bburx=792,bbury=650,width=7cm}
\caption{
\noindent
Distribution of photon pairs as a function of $Q_i$,
with a transverse momentum cut $p_T>$1 GeV.
Pairs of early photons (dottel line), late photons (dashed line)
and total (full line). 
}
\label{FPP03}
\end{figure}

The acceptance is of twofold importance: first, the 
statistics increases with the acceptance, and second--more
important--acceptance determines the maximal range of
relative momenta of particle pairs in correlation analysis.
If the acceptance were too narrow, a measured 
correlation function would not reach its ``plateau"
(the flat region outside the correlation peak), and
the correlation measurement would not be possible. This is 
of particular importance in the proposed
photon correlation measurement, because we are interested
in the correlation intensity at a very large 
longitudinal relative momentum, up to around 2 GeV.
For photons of 1 GeV momentum the acceptance
in the polar angle 
should then be around $\pm$ 45 degrees.
The dedicated photon detector in ALICE, the Photon Spectrometer
(PHOS) (a matrix consisting of lead-tungstenite
crystals) is too short since it covers less than 
a half of the needed acceptance, 
but fortunately, one can apply another photon 
measurement technique, namely the photon conversion technique,
which has been used in modern satellite borne
cosmic gamma ray experiments. 
In ALICE, photons will convert into electron--positron pairs 
in the material of the inner tracker system,
inner field cage of the Time Projection Chamber (TPC),
and in the gas inside the TPC. The created electrons and
positrons may be tracked through the TPC, and the 
information on the original converted photon will be 
obtained from the electron-positron pair reconstruction.
The TPC acceptance for electron-positron pairs 
is sufficiently wide, and 
a good resolving power for
the two tracks of a conversion pair is confirmed in 
the simulation.

The integral amount of material contributing to 
photon conversion will lead to about 5-10\%
conversion efficiency, and we shall assume
5\% overall detection probability.

\section{Results}

It remains to be seen under which conditions the 
HBT correlations of the early direct photons
will be observable, specifically 
in the ALICE experiment at LHC. The main problem is
the presence of photons emitted later
than the early photons, from all other photon sources,
particularly from the decays of neutral pions. 
The HBT correlation at a large relative momentum --
the signature of the presence of the 
early direct photon component -- will be therefore
strongly suppressed.

For simplicity,
in the simulations presented
we assume only 
two photon components: the ``early photons",
emerging at time 0.3 fm/c, from the rapidly developing
parton cascade at
750 MeV effective
temperature, and the ``late photons", comprising 
the later direct photons and the photons emerging
from the decays of neutral pions
(in this simplified approach the late direct photons 
are not separately simulated, but are
effectively assigned as a part of the pion-decay component).
We assume further that the 
fraction of the early direct photons is 5\% of the 
total photon yield.

\begin{figure}[htbp]
\epsfig{file=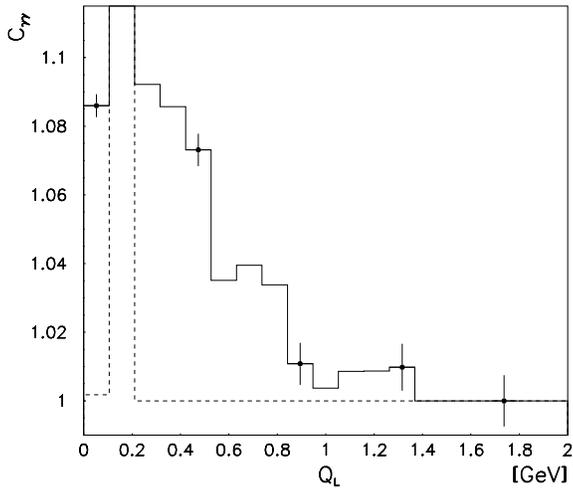,bbllx=0,bblly=0,bburx=792,bbury=580,width=7cm}
\caption{
\noindent
Two--photon correlation function as a function of $Q_L$, with
transverse momentum cut $p_T>$2 GeV. Full line: $R_L$=0.3 fm, $R_s$=3 fm and $R_o$=3 fm;
the correlation enhancement is due to the presence of the 
early photon component.
Dashed line: the early photon component
is shifted to later times by setting $R_L$=6 fm; 
no correlation enhancement is seen.
The $\pi ^0$ decay peak is also visible.
The presented error bars are rescaled and correspond to 
the statistics expected in the experiment, and not
to the actual low statistics used in the simulation.
}
\label{FPP04}
\end{figure}

Assuming a total number of 5000 neutral pions 
per Pb+Pb central collision in the acceptance
of two rapidity units, one arrives at 10,000 
photons from pion decays, 
and 500 early photons 
per event.
Assuming further a detection probability of 5\%,
one would detect 500 late
photons and 25 early photons
per event.

Fig.~\ref{FPP01} shows the transverse momentum spectrum
of the early direct photons, the late photons, and their
sum. The high-temperature early component is strongly 
suppressed at low transverse momenta, below 1 GeV.
Fig.~\ref{FPP02} shows the distribution of photon pairs 
as a function of the
invariant momentum difference
$Q_{i}^2=-(p_{1}^{\mu}-p_{2}^{\mu})(p_{1\mu}-p_{2\mu})$,
where $p_1$ and $p_2$ are the four-momenta
of the two photons.
The contributions shown in Fig.~\ref{FPP02}
are by early and late photon pairs,
while 
the presented total contribution contains also the
mixed early-late photon pairs.
The interesting early direct photon
component is negligibly small compared to the rest,
and in this situation there is no chance to 
observe the early photons' correlations.
The natural way to proceed is therefore to ``enhance" the 
early direct component by applying a high transverse
momentum cut, as suggested by Fig.~\ref{FPP01}.
Fig.~\ref{FPP03} shows that already a cut at 1 GeV indeed 
leads to a significant improvement. In the example presented
below we applied a yet ``stronger" cut, at 2 GeV. 

Two-photon HBT correlations have been simulated 
with the 
longitudinal effective length of 0.3 fm, which
corresponds to the early appearance of the
hypothetical high effective temperature parton gas.
The transverse r.m.s. effective radius was set to
3 fm, which corresponds to the r.m.s. transverse
size of the colliding Pb nuclei. 

Gaussian shape has been assumed for the correlation function 
in all three dimensions, with an intercept of 1.5
(taking into account~\cite{Slotta} 
that photons are bosons of spin 1).
In the results presented we have used an ideal
uncorrelated background. 

Fig.~\ref{FPP04} shows the 3-dimensional correlation function
projected onto the $Q_L$ axis, with a cut 
$Q_S<0.1$GeV and $Q_O<0.1$GeV
(to reduce "dilution"
of the correlation effect). 
A transverse momentum cut at
2 GeV has been applied. 
Only 300,000 events have been generated for this analysis, 
but the error bars presented in Fig.~\ref{FPP04} 
are rescaled to correspond approximately to the expected full
statistics of $10^7$ events!
The expected wide HBT correlation in $Q_L$ due to early photons
is indeed seen. 
In this measurement scheme 
the correlation intensity at large $Q_L$
is the measure of the yield of the early photon component.
The correlation at large $Q_L$ is missing when the early emission
time is replaced by a late one (6 fm/c) in the simulation, 
as shown in Fig.~\ref{FPP04}, dashed line.

\section*{Summary}
A method is proposed to test the existence of rapidly developing
parton cascades in ultrarelativistic nuclear collisions
at RHIC and LHC. 
An early onset of parton cascades in the collision process
should be reflected in abundant emission of photons
which are mutually HBT correlated over a wide
longitudinal momentum scale.
This method
provides the necessary
selectivity for the early emitted photons, since
the late 
photons 
correlate only
over narrower longitudinal momentum scales. 
Based on a rather simple approach,
our study
indicates that
such a measurement could be feasible
in the ALICE experiment at LHC. 
The next step is to replace
our simplistic assumptions 
by input from different detailed models.

\section*{Acknowledgments}
I would like to thank Gordon Baym,
Vesa Ruuskanen, J\"urgen Schukraft, Josef Sollfrank,
Boris Toma\v sik and Urs Wiedemann
for interesting discussions.

\end{document}